***Stumbling around uncharted regulatory structures:***

***NAcrins, or the perspective of specialized sources of modulatory non-coding RNAs***

By

*Marouane Benzaki*

# Abstract and keywords

**Abstract:** The revelation of the supreme authority of nucleic acids in the cellular landscape has precipitated the recognition of the versatility of RNAs in cells. The subsequent discovery of non-coding RNAs was a major breakthrough that revealed their extensive involvement in virtually all physiological processes within the cell. Beyond the barriers of the cell, the current perception seems to support the idea of their participation in intercellular regulation and cross-kingdom communication. However, the presence of non-coding RNAs in the extracellular environment remains essentially a mystery, and the understanding of the significance and the processes governing this presence faces several constraints. This has led us to forge an original and predictive idea that seems to allow an emancipation from the various constraints posed in the current perception of the cited phenomena. In this paper, we will attempt to explore the extent of the probable existence of cellular organizations specializing in the production and management of non-coding RNAs. We will try, through the development of this hypothesis, to draw a picture explaining the significance and logistics of extracellular non-coding RNAs, with an emphasis on microRNAs. This exercise will be realized while relying on and confronting purely theoretical points of view, as well as relevant experimental results. In this manuscript, we will address the presumed morphology, intracellular organization, selective export, transport, transfer, distribution, reception and intracellular function of non-coding RNAs, in the perspective of a regulation cycle orchestrated by "NAcrins" under normal or disturbed physiological contexts.

# Table of Contents

# Acronyms and abbreviations

AGO : Argonaute

EV : Extracellular Vesicle

GI : GastroIntestinal

HDL : High Density Lipoprotein

LDL : Low Density Lipoprotein

RNA : RiboNucleicAcid

mRNA : messenger RNA

ncRNA : non-coding RNA

miRNA : microRNA

naRceptor : Nucleic Acid Receptor

NAcrin : Nucleic Acid secreting specialized source

nt : nucleotide

RNP : RiboNucleoProtein

hnRNP : heterogeneous nuclear RNP

mRNP : messenger RNP

RNase : RiboNuclease

RISC : RNA Induced Silencing Complex

RBC : Red Blood Cell

rER : rough Endoplasmic Reticulum

P-bodies : Processing bodies

SR-B1 : Scavenger Receptor class B type 1

TLR : Toll-Like Receptor

XenoN : Diet-derived exogenous Nucleic acid

# Key points

- Non-coding RNAs and miRNAs in particular appear to be perfect candidates to mediate homeostasis through intercellular regulation, as they may serve as hormones and act as endocrine, paracrine, exocrine and autocrine mediators.

- We put forward an original predictive idea about a ncRNA-centered, specialized cellular organization. The hypothesis seems to overcome the constraints in the current knowledge of intercellular regulation by ncRNAs.

- ncRNAs likely contribute to morphological complexity.

- Different ncRNA secretion pathways and carriers may suggest the existence of transmission channels for special cells with distinct logistical capacities for secretion.

- Cross-kingdom communication involving ncRNAs can be a therapeutic means as well as an epidemiological risk factor.

- Certain tissues and cell types appear likely to be part of a network interconnected by ncRNAs.

# Introduction

The important communitarian reforms in the agrarian world of the 18[th] century brought the world to an unprecedented transition and demographic evolution. The demand for goods and services exploded and, with it, the foundation for the industrial era and technological progress was laid. The energy to run increasingly efficient machines to satisfy the exponentially growing demography accelerated the search for new energy sources. The discovery of petroleum oils and their very high energy density provided an intense stimulus to chemistry, which then worked to prepare this resource for all industrial uses of the time. The technological breakthroughs brought about by the countless scientific innovations and the sufficient maturation of chemistry made it possible to decipher the hereditary information carrier, with the discovery of the structure of DNA as the most notable advance [1]. A few years later, with the discovery of RNA [2], the process of translating genetic information into proteins became clear. This "intermediate" element led to the discovery of similar molecular species, which then led to a cluster of RNA discoveries that provided a clearer, more complete picture of the numerous and diversified physiological roles of these molecules, which are considered essential for homeostasis.

However, the focus on RNAs was almost exclusively on coding RNAs, with non-coding RNAs species being considered only as degradation products of analytical manipulations. It is only very recently – thanks to significant technological advances in sequencing and computational analysis – that the potential of ncRNAs to be important and active molecules in the cell landscape has begun to be considered. The discovery of the first small RNA [3, 4] and its physiological regulatory role in the nematode *Caenorhabditis elegans* paved the way for the discovery of the world of small ncRNAs, which have been shown later to be involved in virtually all physiological processes, in all kingdoms of life. It then became essential to elucidate their biogenesis and mechanisms of action in order to improve our understanding and draw a picture worthy of representing the reality in the field.

Nowadays, it is legitimate to recognize that, when it comes to the classification of ncRNAs in eukaryotes, current knowledge is limited to the "canonical" and "noncanonical" molecular pathways, the latter being poorly studied and underexamined. Thus, there are probably still species of ncRNAs whose biogenesis and functions are not yet known, making a classification even more complex. However, among the best studied classes of these molecules are microRNAs (miRNAs) and Long non-coding RNAs (LncRNAs). While the biogenesis of these two categories is similar in-situ, it differs in the processing as LncRNAs appear to follow the canonical pathways shared with mRNAs processing whereas miRNAs follow additional specific processing steps where they are first synthesized among longer sequences by polymerase complexes, and then generated from characteristic hairpin structures, in a series of successive steps performed by two RNases, Drosha and Dicer [See review in 5, and references therein]. LncRNAs are defined nowadays as transcripts with a minimum length of 200nt with significant disparities in length, binding partners and functional abilities[6], while miRNAs range in length from 19 to 24 nt, and bind to proteins of the Argonautes family. The latter form the core component of RNA-induced silencing complexes (RISC), which recognize specific sequences (or binding sites) of mRNA targets, and thus act as silencing post-transcriptional regulators in the most commonly reported classical scheme [See review in 7, and references therein]. This regulation is reported to be so extensive that it has been estimated that more than 60% of the coding genes have been subjected to the pressure of natural selection to conserve at least 1 site for miRNA binding, and are therefore subjected to miRNA regulation at the post-transcriptional level [8]. LncRNAs appear on the other hand to be paramount to specific context-dependent function as they exhibit relaxed and modulable structures essential to meet the needs of emerging and established complex interactions such as mobilization, scaffolding or topological and spatial configuration of chromatin. This supports literature describing miRNAs and LncRNAs as the architects of eukaryotic complexity, attesting to their roles in virtually all physiological processes, from development to mature physiological function and housekeeping.

When we focus on the idea of such extensive and diverse involvement, it becomes very plausible to think that (A) lncRNAs and miRNAs can be used as signaling and intercellular communication molecules, and (B) that there has been an evolution by natural selection, towards mechanisms of control and selective export, transport and reception of Nucleic Acids, and more specifically ncRNAs such as LncRNAs and miRNAs, since their uncontrolled release would cause aberrations in signaling and intercellular communication.

The two assumptions cited (A and B) are supported respectively by the following elements:

   I)     miRNAs and LncRNAs are present in virtually all biological fluids.

   II)    Several studies have demonstrated significant regulatory effects of extracellular miRNAs and LncRNAs (induced or administered) on target cells [9-14].

III) The fact that complex multi-cellular organisms rely "only" on chemical processes mediated by proteins, lipids or low molecular weight soluble factors, and the cascades derived from them, for intercellular and especially inter-tissue communication – the fact – is probably an epistemological obstacle to be overcome in the current systemic approach.

IV) Some molecular pathways for miRNA and LncRNA exocytosis have been demonstrated [10, 15-18].

V) It appears that the profile of extracellular miRNAs is different from the profile of intracellular miRNAs. One of the contributing facts is that the secreted miRNAs are subtracted from the intracellular content, hence the difference between the two profiles [See review in 19, and references therein].

VI) It has been demonstrated and now established that the majority of circulating miRNAs and LncRNAs are associated to ribonucleoprotein (RNP) or other complexes, including RISC, encapsulated in exosomes, microvesicles, HDL, LDL and other transporters [See review in 20, and references therein].

VII) The prospect of fusion of vesicular shuttles containing miRNAs or LncRNAs with target cells does not cause a problem from a theoretical point of view. An argument contrary to this perspective would lack the insight to give it the benefit of a favorable presumption.

VIII) The concept of naRceptors or miRceptors (nucleic acid receptors or miRNA receptors) is emerging with the recent discoveries of the interaction of membrane Toll-Like Receptors (TLRs) with miRNAs [See review in 21, and references therein]. These findings have major implications, particularly in inflammation, infectiology and immunology.

Based on these observations, it seems logical to extend the investigation to other molecular possibilities used by organisms to ensure homeostasis by intercellular regulation. LncRNAs and miRNAs appear to be perfect candidates for this function, since they can perform the same function in the intracellular environment, whether in the producing or the recipient cell. This idea is gaining acceptance in the scientific community, as it is now supported by a large body of literature attesting the physiological and pathophysiological roles of miRNAs and LncRNAs in a variety of fields, markedly in diseases such as cancer, immune diseases, infections and other disorders.

The role of mediator reserved exclusively for "traditional" factors in the current systemic approach is now opened to the possibility of including miRNAs, LncRNAs, as well as other ncRNA categories to a larger extent. These molecules can thus be considered as hormones in the proper sense of the term, as they probably act as endocrine, paracrine, exocrine and autocrine mediators. However, current theories of intercellular signaling and regulation by ncRNAs and, to a second extent, of cross-kingdom communication, face several constraints, mainly due to the lack of information concerning intra- and extracellular logistics, as well as cellular and tissue distribution. That said, there does not seem to be satisfactory explanations for the phenomena involving circulating ncRNAs and the extent of their roles in homeostasis. These information have led us to forge an original predictive idea which appears to allow an emancipation from the many constraints in the current perception of the phenomena cited, and possibly unveil the reality behind the processes of intercellular regulation by ncRNAs. The hypothesis formulated hereafter is based on the premise that the circulation of ncRNAs happens from a secreting source towards a destination or recipient entity: there are therefore unknown cellular levels of organization or known anatomical structures that specialize in some or all of the processes of production, sorting, packaging, export and balance maintaining of specific ncRNAs aimed to accomplish an extracellular modulation task.

The potential discovery of "*NAcrins*", the etymologically conservative name we propose for these cellular organizations, could constitute a revolution in the understanding of the rudders of physiology. In this article, we will attempt to explore the possibility of the original existence of these cellular organizations in metazoans, with a focus on miRNAs. Then, we will examine the extent of this hypothesis. This will be done in detail and in phases, according to what has been reported in the literature, while relying on and confronting purely theoretical principles, as well as relevant experimental conclusions.

# Development

## Morphology and characteristics. A preconception

When speculating on the characteristics of an unknown cell organization, one must be careful as to avoid any interpretation that could deviate the process of mental representation from reality. We will therefore try to adopt a questioning posture in order to briefly explore the subject, while proceeding with an explanatory approach.

The first multicellular forms certainly succeeded in finding a balance between the available supply of resources and the demand for them in different parts of the body. In this context, several molecules would have been perceived by other cell populations in the organism to signal the status, including that of the needs, of the source cell population. These molecular precursors most likely served as the basis for the construction of subcellular and cellular structures that were more efficient in communicating the needs, in a manner that promoted the maintenance of beneficial mutuality among all components of the organism. For example, this is apparently the case of some neuropeptides, which, having appeared as precursors in early forms such as cnidarians, will see their catalog diversify in subsequent complex species that will later develop subcellular and cellular structures specialized in the transmission of neuropeptides, such as nerves, ganglia and brains [22-24].

This is plausibly the case for a number of molecules which, by the mere fact of their presence and according to contextual parameters, have been gradually employed in processes serving a physiological function, and have in some cases considerably influenced the fate of cell populations and organisms that employ them. Indeed, miRNAs or similar gene regulatory ncRNAs are present in virtually all species representing the entire tree of life. This may testify in favor of their early presence in the evolution of living organisms, and therefore attest to their involvement, adaptation and influence in subsequent species lines. In the same context, it has been proposed that an ancestor of eukaryotes possessed a miRNA processing system [25]. However, since few similarities have been detected between the miRNAs of the different branches of eukaryotes, it is considered that the miRNAs evolved independently in each lineage [26]. In metazoans, it appears phylogenetically that several expansion waves of the miRNA repertoire have taken place at the base of the lineage of bilaterians, especially nephrozoa, and at the base of the vertebrates and placental mammals [27-30]. It has also been reported that a preserved miRNA is rarely lost in descending lines [30-32]. In addition, these expansion events appear to correlate with an increase in the morphological complexity of organisms [27, 30, 32, 33]. This seems to lean in favor of the influence of miRNAs as a molecular basis, and as a promotor of the evolution towards diverse and morphologically complex multicellular forms [27, 29, 30, 33]. It is, in this sense, plausible to think that certain subcellular and cellular structures were oriented in favor of a management of these molecules, which would have later led to specialized structures, as assumed in this article. However, since complexity is a subjective notion, one must question its degree at the level of subcellular and cellular organization and the contextual parameters that shaped these structures, probably over a very long evolutionary period. It is also legitimate to wonder about the associated energy costs and the material resources recruited and oriented by these cellular organizations towards a miRNA-centered lifestyle in different contexts. Since the adaptation to produce, select and secrete in a well-controlled manner probably requires compromises and concessions, because of the deviation of a considerable part of the resources towards increasingly specializing subcellular and cellular components, whether it being a matter of randomness or necessity, or both. That said, and on the contrary, complexity in itself may not be necessary, nor necessarily advantageous if the organization in question reaches a certain level allowing it to operate efficiently with the existing means.

**Fig.1: miRNAs acquisition and morphological complexity in metazoan phylogeny**

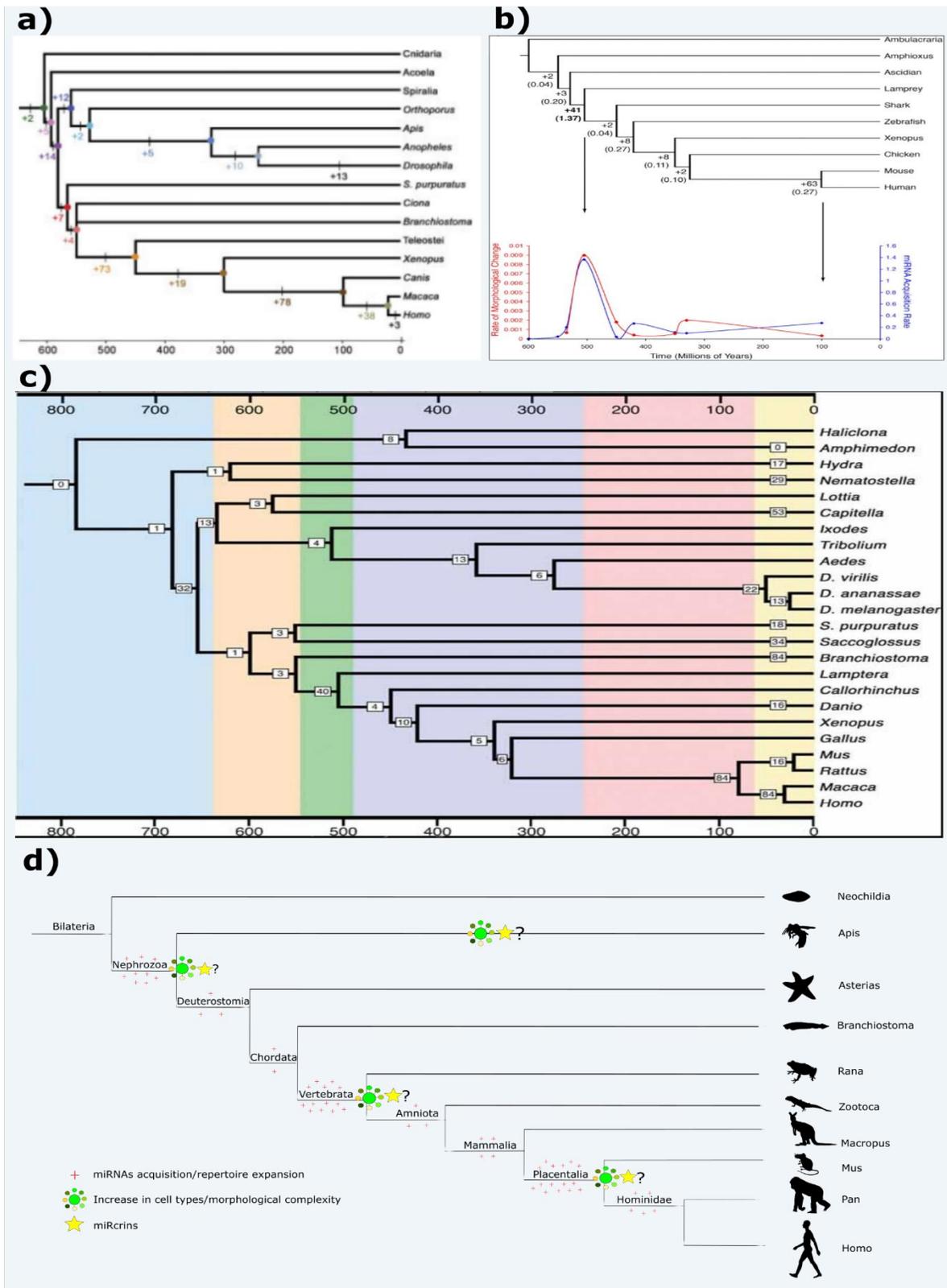

**Fig.1: a)** Phylogenetic tree with the relative changes to the 243 human and the 70 fruit fly non-paralogous miRNAs indicated at each node, (Adapted from *Sempere, L.F., et al., The phylogenetic distribution of metazoan microRNAs: insights into evolutionary complexity and constraint. J Exp Zool B Mol Dev Evol, 2006. 306(6): p. 575-88. ©2006 WILEY-LISS*) [31]. **b)** Evolutionary acquisition history of the 129 chordate-specific families of miRNAs found in eutherian mammals, (Note. Adapted from *Heimberg, A.M., et al., MicroRNAs and the advent of vertebrate morphological complexity. Proc Natl Acad Sci U S A, 2008. 105(8): p. 2946-50 ©2008 PNAS*) [29]. **c)** The acquisition of miRNA gene families from the Cryogenian (light blue), through the Cenozoic (dark yellow) for 24 metazoan taxa, (Note. Adapted from *Peterson, K.J., M.R. Dietrich, and M.A. McPeek, MicroRNAs and metazoan macroevolution: insights into canalization, complexity, and the Cambrian explosion. Bioessays, 2009. 31(7): p. 736-47. ©2009 Wiley Periodicals*) [28]. **d)** The acquisition of miRNAs and the expansion of their repertoire correlates with an increase in cell types and morphological complexity. The most notable expansion events appear to have occurred at the nephrozoa, vertebrates and placental mammals lineages. miRcrins, or more comprehensively NAcrins, may have emerged among the novel cell types and developed as a part of the novel architecture in subsequent lineages.

We are nevertheless tempted to put forward schemes of such organizations based on a function-structure relationship, following the rationale that morphological complexity is mainly due to particular molecules and regulatory elements. One could think of cell aggregates that have reached the level of a morphologically distinct anatomical structure, with different cell types, each capable of supporting a special function, while working in a coordinated manner. These cells would be characterized by adapted subcellular compartments, functioning according to a purpose-centered metabolism. The cells could also perform different physiological functions, and be part of existing systems that have been proportionally influenced by miRNAs and other molecules used during evolution. In this sense, familiar anatomical structures would handle the logistics of miRNAs in addition to the physiological functions for which they are reputed, based on the same molecular mechanisms or other unexplored pathways, and following a shared metabolism.

In addition, it has been proposed that the evolution of certain organs occurs through the acquisition of attributes of existing tissues, through changes in patterns and development, and through the evolution of new cell types [34]. The evolutionary origin of organs seems to be, according to the same authors, often the result of new interactions between distinct cell populations [34].

Although NAcrins can take a form other than that of an organ, it is still logical to consider the recipient or target tissues of miRNAs, and to question the possible interactions between the target cell population and the NAcrin precursor cell population. The evolution of the NAcrins may have been target-dependent, and elucidating feedback loops or an intraorganismal in-vivo leitmotif associated with the miRNAs under study could potentially lead to the discovery of the structures in question.

In short, it is complicated to speculate on the morphological characteristics that could be present in NAcrins, since evolution by natural selection can lead to unusual patterns according to the pressure-imposed paths throughout the period under consideration. NAcrins can thus be in a primitive or advanced evolutionary state in humans, be organized in cellular aggregates or networks with apparent or absent architecture, and be of microscopic or macroscopic anatomical nature. We can think of a single source as we can think of multiple sources, the latter may have similar or different morphologies, and may be scattered in strategic locations for their functions. In this sense, they can take the appearance of glandular structures, organs or tissues, or they can be part of glands, organs or tissues whose anatomy is known, but whose functions concerning the miRNAs are unknown.

## Intracellular logistics and selective export

### Mechanism of action and function of miRNAs

In order to look at the intracellular distribution of miRNAs, one must first consider their reported mechanisms of action. It appears that the silencing of gene expression resulting from miRNAs action follows two pathways that can be distinguished according to the events involved, notably the state of docking for the nucleolytic step. The first is a pathway characterized by repression and obstruction of the transcription and translation of the target mRNAs. In this sense, the mRNA can be either in synthesis, awaiting translation or in an active translation state. Here, the complementarity of miRNA/mRNA sequences is generally partial and arranged in a structurally unstable and inadequate way for the RISC scissors to reach and cleave the target [See review 35, and references therein]. In this case, the recruitment of additional elements will obstruct and inhibit the mRNA target's translation [See review 36, and references therein]. The second pathway is characterized by the cleavage and decomposition of the target mRNA, plausibly in a mature state, thus before, during or after translation. In this case, complementarity is sufficient in the central portion of the base pairing and the docking is structurally appropriate and kinetically stable, allowing the nucleolytic domain of the RISC to reach and cleave the mRNA target [See review 35, and references therein].

# Table.1: Main experimental data supporting the mechanisms of miRNA action

| Proposed mechanism | Main experimental data supporting given mechanism | References | Additional comments |
|---|---|---|---|
| M1. Cap Inhibition (Inhibition of translation initiation via cap-40S association) | 1. IRES-driven or A-capped mRNAs are refractory to microRNA inhibition. 2. Shift toward the light fraction in the polysomal gradient. 3. GW182 involvement in the suppression of initiation via cap-40S association. | Pillai et al. 2005; Humphreys et al. 2005; Kiriakidou et al. 2007; Thermann and Hentze 2007; Filipowicz et al. 2008; Eulalio et al. 2008b; Zipprich et al. 2009; Hendrickson et al. 2009 | Postulated: initiation inhibition upstream of eIF4G recruitment by eIF4E, suppressing the recognition of the cap by eIF4E. |
| M2. 60S Joining Inhibition (Inhibition of translation initiation via 40S-AUG-60S association) | 1. A lower amount of 60S relative to 40S on inhibited mRNAs. 2. Toe-printing experiments show that 40S is positioned on the AUG. | Chendrimada et al. 2007; Wang et al. 2008 | It is important to point out that, strictly speaking, there exists no proof of the effect on the AUG scanning in this work, although some authors (Nissan and Parker 2008) interpret this data as an inhibition of scanning. |
| M3. Inhibition of elongation | 1. Normal polysomal distribution of the inhibited mRNA. 2. Sensitivity to EDTA and puromycin indicating functional, translating polysomes. 3. Some mRNAs can be repressed by a microRNA even when their translation is cap independent (IRES or A-capped mRNAs). 4. Ribosome "stay" longer on the inhibited mRNA. 5. Decrease in the number of associated ribosomes. | Olsen and Ambros 1999; Landthaler et al. 2008; Maroney et al. 2006; Petersen et al. 2006; Lytle et al. 2007; Gu et al. 2009; Baillat and Shiekhattar 2009; Karaa et al. 2009 | It is important to note that it is really difficult to discriminate experimentally between different post-initiation mechanisms (elongation inhibition, ribosome drop-off, or normal elongation with nascent polypeptide degradation). Possibly the polysomal profile should be slightly different, showing a normal profile in the case of nascent protein degradation, fewer ribosomes per mRNA in the case of elongation arrest, and the smallest ribosome number per mRNA in the case of drop-off. |
| M4. Ribosome drop-off (premature termination) | 1. No difference in polysomal profile in presence of miRNA. 2. Addition of puromycin shows actively transcribing polysomes. 3. Any nascent peptide was detected. 4. The read-through codon-stop and more rapid loss of polyribosome upon initiation block. 5. Decrease in the number of associated ribosomes. | Petersen et al. 2006; Wang et al. 2006; Hendrickson et al. 2009 | |
| M5. Cotranslational protein degradation | 1. Sedimentation of the mRNA together with miRNA-RISC complexes in actively translating (puromycin-sensitive) polysomes. 2. Polysomal profile, suggesting that the repressed mRNA is actively transcribed. | Nottrott et al. 2006; Petersen et al. 2006; Pillai et al. 2005; Wang et al. 2006; Maroney et al. 2006; Gu et al. 2009 | 1. No nascent peptide has ever been experimentally demonstrated. Possibly, its degradation occurs extremely rapidly after the synthesis. 2. This degradation, if it exists, was shown to be proteasome independent, but no other specific protease or complex involved in it has ever been identified. |
| M6. Sequestration in P-bodies | 1. In situ hybridization revealed localization of miRNA, mRNA, and RISC complex inside cytoplasmic structures called P-bodies. 2. In P-bodies, translational machinery is absent and degradation machinery is enriched (local concentration of all needed enzymes). | Pillai et al. 2005; Sen and Blau 2005; Jakymiw et al. 2005; Liu et al. 2005a,b; Bhattacharyya et al. 2006; Leung et al. 2006; Pauley et al. 2006; Eulalio et al. 2007a | 1. There are two different propositions about the P-bodies' function: (a) sequestration of targeted mRNA apart from translational machinery; (b) a kinetics advantage for mRNA decay. 2. The main concept today is that P-bodies are not required for but rather a consequence of microRNA-driven translational inhibition. 3. Only a small portion of miRNA, mRNA, and RISC complex is localized inside P-bodies. |
| M7. mRNA decay (degradation, destabilization) | 1. Decay of targeted mRNA occurs without direct cleavage at the binding site. 2. Only a slight protein decrease can be obtained by translational inhibition alone. When the protein level decreases by >33%, mRNA decay is the major component of microRNA-driven silencing. 3. Different details of decay mechanism have been shown: decay by mRNA deadenylation, decapping, or 5' to 3' degradation of the mRNA. | Coller and Parker 2004; Lim et al. 2005; Bagga et al. 2005; Jing et al. 2005; Behm-Ansmant et al. 2006; Wu et al. 2006; Eulalio et al. 2007b; Wakiyama et al. 2007; Filipowicz et al. 2008; Baek et al. 2008; Selbach et al. 2008; Hendrickson et al. 2009; Guo et al. 2010 | 1. Degradation mechanism is usually coupled with translational inhibition. 2. In some studies the translational inhibition had the same efficiency with or without mRNA degradation. 3. Depending on the mRNA, two different cases for mRNA degradation via microRNA are possible: Ongoing translation is required for the decay, or else decay occurs in the absence of active translation (Eulalio et al. 2007b). |
| M8. mRNA cleavage | 1. Full complementarity between microRNA and its mRNA target. 2. RNA fragments diagnostic of directed target mRNA cleavage. 3. Down-regulation of corresponding target mRNA. | Rhoades et al. 2002; Llave et al. 2002; Hutvagner and Zamore 2002; Yekta et al. 2004; Bagga et al. 2005; Valencia-Sanchez et al. 2006; Aleman et al. 2007 | 1. mRNA cleavage occurs only if microRNA is fully or near-fully complementary to its target. 2. It is similar to siRNA-mediated mRNA cleavage mechanism. 3. mRNA cleavage was proved to be very common for plants, and much rarer in animals. |
| M9. Transcriptional Inhibition (miRNA-mediated chromatin reorganization following by gene silencing) | 1. Complementarity between some microRNAs and promoter sequences of target genes. 2. microRNA increases methylation of the targeted mRNA promoters. 3. Evidence for direct nuclear microRNA import. 4. The levels of target RNA transcripts are strongly reduced, while no mRNA decay is detected. | Kim et al. 2008; Khraiwesh et al. 2010 | 1. siRNA-mediated transcriptional repression was shown by Morris et al. (2004) and Weinberg et al. (2006). 2. The possibility of miRNA-mediated transcriptional activation was also shown (Place et al. 2008). |

**Table.1:** Adapted from *Morozova, N., et al., Kinetic signatures of microRNA modes of action.*

It is important to note that 3 of the 4 human endonucleases reported to be fundamental to RISC, namely Argonautes, are catalytically inactive [35]; only binding to AGO2 is likely to allow cleavage, provided that the seed sequences correspond correctly, in addition to an auxiliary but unnecessary support of secondary complementarities [35]. In this sense, it seems that mammals rely more on the first mode of action [35]. It is reasonable to lean towards similar conclusions because of the low catalytic power and the almost exclusive partial complementarities for target regulation. According to some authors, miRNAs seem to act, in most cases, as rheostats allowing precise adjustment of protein production [38]. Some authors even go so far as to stipulate that the role of miRNAs is above all to reduce noise expression instead of serving as real mediators [39-41]. It is true that ncRNAs are a new field and it is customary to try to explain many unknown phenomena by the simple existence of the novelty. miRNAs may have a limited role, which is controlling expression noise, and that the conclusions reported in the different fields concerning them are the result of this role only. However, it is considered more accurate to retain what applies to one or a few cell types and contexts. An initial interpretation based on all these elements would first require that we consider that miRNAs have a very large number of potential targets, some of which are likely to cause their platform to undergo radical changes in expression, and others less so. In the latter case, miRNAs can act on several mRNAs, and conversely, several miRNAs can act on a single mRNA.

Now, assuming that the role of miRNAs is largely centered on mRNAs, the locations visited by an mRNA during its lifetime can logically be considered the most likely locations of miRNAs and the machinery associated with silencing. In this sense, apart from the obvious location of initial synthesis, several authors have reported proven and potential sites of processing and translation of mRNAs, with more or less proven colocalization of miRNAs and the enzyme complexes responsible for silencing. These include the nucleus, rER, Golgi complex, P-bodies, endosomes, granules, mitochondria and other sites [See review 42, and references therein]. It is here that it becomes evident that the framework of miRNA activities in the different sub-cellular spaces, especially at the P-bodies level, is a highly debated topic because of the great functional diversity they demonstrate and the methodological challenge posed by detection and characterization. Thus, the subcellular localization of miRNAs provides information on their implications in a surprising variety of processes, such as adaptation to cellular conditions or events, stress response, phenotypic changes and differentiation at the nuclear level, mitochondrial metabolism, etc. [See review 43, and references therein]. In this sense, it is likely that, depending on the cell type, a batch of special endogenous or exogenous miRNAs, selected and guided by RNPs or associated factors, is intended to target each compartment or key step.

## Selection and secretion of miRNAs

One can think that "functional" miRNAs can be selected by their source in an active way in order to induce intra- or extracellular regulation. In the latter case, and although no selection per se has been demonstrated thus far, there is increasing evidence to suggest that this process takes place in a pattern leading to the secretion of miRNAs in the extracellular space, thereby leading us to think in terms of selective secretion. In the same vein, it appears that miRNA secretion occurs in exosomes following a ceramide-dependent pathway [10]. The cells in question in this study, whose nephrotic origin is important to note here, exhibit a ceramide-dependent secretory machinery, the biosynthesis of which is regulated by nSMase2, allowing the release of exosomes carrying miRNAs that have been demonstrated to silence their target gene in recipient cells. Other teams have succeeded in demonstrating the involvement of other secretion pathways, namely SNARE transmembrane proteins for exocytosis in neuroendocrine cells [17], as well as a non-vesicular/AGO2-linked release [16]. However, the intracellular organization and the compartments responsible for secretion remain to be fully elucidated. It is perceptible that the traffic of miRNAs orbits, according to the current knowledge, around structures such as P-bodies and GW-bodies, among others, and it is moreover accepted that these sites are rather a consequence of silencing. These structures appear to be closely related to elements having direct or indirect contact with miRNAs and ncRNAs. Some authors, therefore, speculated on a sorting based on associations with GW182 [44], a protein family of the RISC complex. Other authors suggested a sorting based on localization signals on particular associated mRNAs, on mRNPs associated with motor proteins, or on the pre-miRNA [See 45, and references therein]. Still others suggest mechanisms of sorting based on hnRNPs, proteins with an RNA-binding domain and some membrane proteins [See 46, and references therein]. On the contrary, other teams have demonstrated the presence of hexanucleotides serving as a transferable signal for nuclear localization, which are therefore used for import into the nucleus [47]. Although miRNAs are found in all cell compartments, their localization most likely depends on the cell type and, as a result, signals for localization and sorting for the purpose of export may deploy only in cells specialized for this task, or cells with attributes of the selective secretion necessary for the proper functioning of their environment. In this case, it would be logical to think that these specialized cells promote the heterogeneity of miRNAs between intra- and extracellular media and, consequently, cause the appearance of signatures in their environment or in the target environment. In fact, it has been reported that significant differences between intracellular and extracellular miRNAs profiles exist [See review 19, and references therein], and that some tissues have special signatures of miRNAs [48-50]. These signatures may be a simple consequence of the heterogeneous expression of miRNAs depending on the genetic configuration of the tissue or the likely consequence of specialized sources. However, it remains complicated from a methodological point of view to determine, after extraction of the miRNAs, their origin in relation to what has been discussed.

On the contrary of these considerations, one can consider a passive selection on the target side. Here, we can evoke the concept that several low affinity miRNAs can help by buffering the many potential targets of a specific miRNA, thereby acting to increase the regulatory efficiency of the latter, whether endogenous or exogenous. Following the same logic, different cell types can adopt configurations that allow the identification of a gene or a group of genes for regulation in a specific environmental setting, which might also explain the tissue signatures previously discussed. This mechanism can be considered as a passive selection mechanism allowing a specialized source, through selected or non-selected "functional" miRNAs, to act in a specific regulatory way on their preferred target. This concept may also prove to be extremely complicated to elucidate from a methodological point of view.

Besides, the demonstrations cited, even if they were performed in an in-vitro context and in atypical cells from different sources, can only reinforce the idea that such an investment in energy and adaptation of complex and highly conserved machinery to convey miRNAs, is an indication of the high probability that there is a selective secretion of miRNAs intended to accomplish an extracellular task, for which they were produced, by internal or external stimuli, selected, and conveyed for secretion. It also appears that these extracellular miRNAs are either released in a free state or transported by different carriers, which necessarily implies the existence of different secretion pathways. The remaining and most important challenges are, therefore, to elucidate the existence of unexplored mechanisms for labelling, sorting and secretion of miRNAs intended for export in different cell types, and to determine the contributions of different organs and tissues to the extracellular content of miRNAs. These challenges can be proposed within an approach that considers cell types that have developed during and after major evolutionary events involving miRNAs, such as the considerable expansions of their repertoire in relation to the increase in morphological complexity, for example.

## Carriers, stoichiometry and transfer

The existence of miRNAs outside the cells has long been perceived as a consequence of cell death or perturbations, especially in pathophysiological contexts, such as cancer. Demonstrations of secretion, as well as suggestions of the consistency of serum miRNA levels in healthy individuals and of the stability by resistance to extracellular conditions [51] have paved the way for a wide range of discoveries concerning extracellular miRNAs.

Indeed, it is now accepted that miRNAs circulate in all biological fluids while remaining in stable form, using "shuttles" that have, among other functions, a role of protection against degradation in these different media. It even appears that there are variations in the concentration and profile of miRNAs between different biological fluids analyzed [52]. Extracellular miRNAs have been reported to be linked to RNP complexes, including AGO2, and circulating freely in biological fluids outside of vesicles. They are also loaded into vesicles, mainly exosomes and microparticles, or are transported by lipoproteins (HDL, LDL) [See review 53, and references therein]. It is surprising, at first, to observe this variety of reported carriers. And it is all the more intriguing when we come down to the idea that we do not know the origin and functions of these extracellular miRNAs, a consequence that can be attributed to the numerous and complex methodological challenges.

## Vesicle-free miRNAs

It has been suggested that a large proportion of plasma extracellular miRNAs are of non-vesicular origin and circulate in association with Argonautes [54]. The same authors suggested that these non-vesicular plasma miRNAs are mostly the result of cell damage and remain in suspension due to their high stability, despite the highly labile nature of RNA. However, it remains consistent to consider the possibility of selective secretion for the purpose of regulation. In this sense, it has been shown that plasma miRNAs of non-vesicular origin allow the regulation of immune cell function [55], which increases the probability of a selective secretion or, at least, of a selective internalization. In the latter case, the specificity of the interactions remains to be demonstrated. Although the community remains skeptical about the possibility of a physiological function of non-encapsulated extracellular miRNAs, it is conceivable that selection processes may exist in the context of their uptake by the liver or the kidneys or their use by cardiovascular or immune cells for example, either locally or systemically. Apart from the conventional mode of action of miRNAs, they can act as specific ligands and mediate a physiological response by triggering signaling cascades. In this sense, miRNAs have been shown to bind to TLRs and induce an inflammatory response [56]. The existence of naRceptors, or miRceptors capable of recognizing miRNAs in the context of a paracrine or endocrine interaction has been proposed [57]. Although it is plausible that certain non-vesicular extracellular miRNAs may be secreted from particular cellular subsets, the remaining challenge is to determine their source in-vivo and the specificity of the interactions by which these molecules act in the context of an unconventional mode of action.

Encapsulated miRNAs and stoichiometric considerations

Concerning miRNAs encapsulated in extracellular vesicles (EVs), it is interesting to note that the profiles of miRNAs are different according to the class of vesicular transporters [58]. Although there are sub-populations for some classes of EVs, it is important to recognize that there are no exact boundaries between these different sub-populations and that current sorting methods are not sufficiently discriminating. This fact is limiting, since each may have unique characteristics in terms of source, mode of secretion and uptake, and therefore the miRNA content of each one cannot be accurately determined since most analyses are based on heterogeneous populations of EVs.

> **Box 1**
>
> EVs encompass a large population of bilayer vesicles secreted by cells in the extracellular space. These vesicles have recently been recognized as functional cargo transporters capable of modulating recipient cells. They are thus considered an essential form of intercellular communication. This potential has therefore attracted the attention of the scientific community which has begun to appreciate the different types of EVs and their functions. The term EV is a generic term for a heterogeneous population of vesicles, which can be divided into subtypes according to their biogenesis, method of secretion, size and function. Although there are no strict delimitations based on these parameters, broadly 3 subtypes can be distinguished. (i) Exosomes constitute the first subtype and are formed in the endosomal pathway before being released into the extracellular medium by fusion of the multi-vesicular bodies with the cytoplasmic membrane. Their size ranges between 40 and 120nm. (ii) Microvesicles constitute the second subtype and are formed by direct budding of the cytoplasmic membrane. Their size ranges between 100nm and 1μm. (iii) Apoptotic bodies are the third and last subtype and are formed following apoptosis, their size ranging between 50nm and 2μm.
>
> Lipoproteins are monolayer spherical particles. They are assemblies of lipids and proteins that form in the liver and intestine. Recent discoveries concerning the associations of lipoproteins and miRNAs have renewed the interest of the scientific community in their potential as functional cargo carriers. Lipoproteins are traditionally classified by size and density. They are most often classified into the 5 categories of HDL, LDL, IDL, VLDL and chylomicrons. Their size ranges from 5 to 80nm.

This element has often led the community to consider the quantity of miRNAs transported by EVs, failing to be able to demonstrate the existence of EVs enriched with a load made up of selected elements. It is obvious of course that the number of miRNAs transported varies according to the carrier. Considering the volume and molecular weight of a miRNA, and the fact that it is most often associated with RNPs in complexes with a mass of up to 300 kDa, according to some characterization studies [54], EVs, such as exosomes or microvesicles, can carry a large number of them, among other transported molecules, due to their large size. Lipoproteins, being the smallest particles, are likely to carry 10,000 copies per μg of HDL [59].

For this reason, exosomes have been among the most studied EV subtype because of their sufficient size and the results of investigations demonstrating the incorporation of different molecular loads, notably miRNAs. On the receptor side, the number of molecular targets is very large if we take into account the functional diversity of miRNAs, and if we go beyond the miRNA 1:1 mRNA stoichiometry oversimplification. In this case, the number of mRNA copies produced per gene becomes a critical data in order to evaluate the necessary load of the EV to induce a physiological effect in the recipient cell. It is certain that each cell type has its own repertoire, and, more specifically, each gene has a certain "pace", controlled at two levels, transcription and translation. For reasons of energetics and precision, some genes lean towards different transcription/translation ratios to achieve the same number of proteins. As an illustration, if we assume that the energy cost of an RNA is higher than that of a protein, a gene, depending on its function, would produce a single mRNA at $T_0$, from which 100 proteins will be translated. There would therefore be an energy saving but an unprecise quantity at $T_1$ due to translation fluctuations (uncontrolled delays due to nuisance for example). Other "vital" genes that are subject to a lot of potential nuisance reduce these unexpected fluctuations by producing 10 mRNAs at $T_0$, each translated into 10 proteins. There would therefore be less cost savings for more precision in quantity at $T_1$. The difference in mRNA copy number can therefore be very large, not only because of what has been mentioned, but also because of the conditions that the cell might be exposed to. However, it seems that the number of mRNAs produced per gene is on average 2 molecules per hour [60], an approximation to be used with caution with regard to what has been mentioned. It is therefore conceivable that the miRNA cargo of the carriers is likely to be sufficient to exert a regulatory effect after their arrival in the recipient cell. However, studies of miRNA and exosome stoichiometry suggest that an exosome may contain, on average, far fewer than a single miRNA molecule [61]. Other authors, on the contrary, suggest exosomes are enriched in particular miRNAs, with a number of miRNAs in at least 10 copies per exosome [62]. The first authors, nevertheless, recognized through their proposed models that there could be some enriched exosomes containing a high concentration of selected miRNAs, this being the so-called high concentration/low occupancy model [61]. This brings us back to the important consideration of the different EVs sub-populations and their loadings, that can be enriched with selected miRNAs, probably by cell types specialized in the export of miRNAs for modulatory use.

Lipoproteins, although smaller, also appear to be able to carry sufficient copies of miRNAs to induce a physiological response. The export of HDL containing miRNAs, which have been shown to regulate mRNA targets, suggest the involvement of these particles in intercellular communication by miRNAs [58]. On the other hand, although most studies have focused on HDL, LDL, which appear to carry a distinct miRNA group, also seem likely to carry enough miRNA molecules [59].

Beyond capacity, it is also noteworthy that the carriers carry distinct loads, which would suggest the existence of a transmission spectrum. Together, these deductions may indicate that there is a probability that the classes and subclasses of miRNA carriers are numerous with respect to the variation in signatures by carrier class, and the likelihood of the existence of subclasses with high concentrations of selected miRNAs. In this sense, we are tempted to suppose an interdependence between transporters and cellular sources, the latter having sorting and secretion mechanisms oriented towards different targets. The existence of NAcrins, specialized in regulating a compartment through an exclusive transporter, is a reasonable explanation of the variety of transporters and associated signatures observed, and, at the same time, a possible step forward towards the elucidation of the physiological functions of extracellular ncRNAs.

## miRNAs transfer

Assuming the existence of carriers carrying specific cargos and that of miRNAs in all branches of the tree of life, it seems plausible to think that there could be shared regulatory mechanisms through miRNAs, and that a transfer of specific XenomiRs and to a larger extent Xeno Nucleic acids (XenoNs) to a metazoan consumer is a potential therapeutic means. In this case, we could speak of cross-kingdom communication and regulation. Several teams have based their work on this possibility in order to develop therapeutic approaches. It is even possible to imagine explanations, without pretending to ignore the biases, for the ethnopharmacological principles that have formed the basis of certain alternative medicines through phytotherapy and unconventional foods. It becomes important here to question the crossing of the Gastro-Intestinal (GI) barrier. Some authors proceeded to feed mice with rice particularly rich in a specific miRNA, and detected the presence of the miRNA in question in the sera and tissues of the animals, implying a possible passage from the GI tract to the blood and organs [63]. However, these results and the results of other similar studies have been highly debated, since several authors have been unable to replicate them. That said, the transfer of miRNAs from food to blood remains inconclusive [See review 64, and references therein]. However, a passage through the various barriers of oral intake, in particular the GI tract, is theoretically possible. Indeed, the GI tract contains a wide range of different cells and has a very large surface area, which makes it susceptible for the reception and absorption of miRNAs. Carriers play an extremely important role here, since they can influence the success of the passage of the different barriers. Furthermore, miRNAs have also been shown to remain stable when transported in exosomes and to resist degradation and denaturation under different conditions, including those present before and during the oral route, such as RNases, high cooking temperature and gastric acidity [65, 66]. That said, if it becomes apparent that the transfer of a certain amount of xenomiRs is possible, then pharmacodynamics, pharmacokinetics and other key pharmacological parameters will require further investigation. Some authors have even put forward the hypothesis that dietary xenomiRs contribute to the content of miRNAs in an organism, and that these xenomiRs influence the homeostasis of miRNAs [67]. These same authors rely on HDL as a transporter, and suggest that presumed biomarkers of certain diseases would rather indicate a response to a qualitative or quantitative dietary alteration by xenomiRs. The comments put forward by these same authors [67] seem to support the NAcrins hypothesis. In this sense, we envision that NAcrins maintain an equilibrium of miRNAs and ncRNAs in certain compartments or systemically. In our model, the intake of specific xenomiRs/vehicle-rich foods could rather stimulate the mechanisms in place for intercellular regulation and restore or accentuate differences in concentrations of specific miRNAs. This would, therefore, cause a disruption in the balance of ncRNAs maintained by the NAcrins and induce pathological manifestations or restore the ncRNA balance and induce a beneficial therapeutic effect.

### Recipient cells, targets and tissue distribution

All cells in the body can obviously internalize extracellular miRNAs. However, we will briefly outline some examples that appear likely to be part of a network interconnected by miRNAs.

In relation to the previous section, one could first think of certain cells of the GI tract that can be considered both targets of xenomiRs and a "mandatory" source of extracellular endogenous miRNAs. In this case, it is possible to envisage that certain cell types of the GI tract are involved in the capture of xenomiRs by simple transporter fusion or by endocytosis mediated by surface markers. These molecules would be internalized and, in the same illustration, the cells could serve as a sorting barrier to eliminate foreign miRNAs and redirect some to the appropriate transmission channels, in order to regulate or precondition surrounding or trans-tissue targets. These cells may have a transistor-like ability to produce functional miRNAs, coupled with a logistical capability to sort miRNAs, endogenous or exogenous, for secretion into the appropriate transmission channels. We thus speak of a cellular regulatory complex capable of acting as a filter and as a distribution center located between the exogenous supply and the endogenous equilibrium in extracellular miRNAs. This use could be explained by the permanent contact of these cells with exogenous resources and the microbiome, through a range of molecules, some of which are potentially hazardous to the host, hence the need to maintain monitoring and control over signals. The host organism has indeed been shown to shape the intestinal microbiome via fecal miRNAs [68]. The authors of this study demonstrated the contribution of epithelial cells, and some specialized cells, such as Paneth and Goblet cells in the modulation of the microbiome through specific miRNAs. There may have been, during the evolution of this tissue, a reorientation of certain cells to communicate with internal cells.

The GI tract may also serve as a simple passage without having any special role, in which case some xenomiRs would induce an effect on the receptor cells in the GI tract, and the rest would be recycled, or would cross through transcytosis. The GI tract may also be a direct target of extracellular miRNAs originating from an internal source and, with regards to direct targets, it is important to consider the cellular recognition mechanisms of the vehicles. To the best of our knowledge, absorption at the level of target cells remains poorly studied and based on hypothetical views. As a result, it has been proposed that the absorption of vesicular miRNAs is accomplished via macropinocytosis or clathrin-mediated endocytosis [69]. Still, others have proposed SR-B1 as a receptor for the admission of miRNAs associated with HDL [58]. It is also possible to foresee simple fusion, direct uptake of vesicle-free miRNAs, or direct transfers between cells through gap junctions.

Reception can also be carried out in several stages, since the notion of target can rationally include transit targets. These forwarding agents could be a kind of reservoir that could serve as outbound logistics operators. They may be located in a given location, or they may be mobile, allowing the miRNAs to be routed to their final target. Here, one can think of circulating blood platelets and Red Blood Cells (RBCs), among others, which are known to contain significant quantities of miRNAs that can be transmitted through EV release following their stimulation [70, 71]. Both papers emphasize the importance of these entities in inducing functional changes in their different recipient cell types. In this sense, it is theoretically possible that anucleated platelets can collect miRNAs [72], as they pass through loading compartments, or plausibly acquire a repertoire of miRNAs as they mature, and then deliver the part corresponding to the received signal upon arrival at the destination. The cardiovascular system can be envisioned through this lens as a prominent body with intricate diversity and assemblage capable of supporting all or part of the transmission and logistical spectrums, as well of course as the full biological cycle from secretion to uptake in its array of tissues, organs and cells. The cardiovascular system and its central role in overcoming the passive diffusion limitations that hindered the development of larger and more complex organisms may have pushed for and/or developed in tandem with NAcrins as these organisms faced challenges that required major adaptations to accommodate for different environments and lifestyles, bringing the cardiovascular system as the potential main NAcrins circuitry and circulation platform of ncRNAs.

Since we have come so far, why not touch on the myth of Prometheus to remember, at its true value, the supreme role of the liver. Some people think that we had, so long ago, some knowledge of the regenerative capacity of the liver. Could this be one of the reasons for the appearance of the myth, the hepatocentric doctrine and the interest of several civilizations in this organ? Perhaps, on the contrary, the liver was chosen to allow a recurrent punishment of the soul of Prometheus - the liver being its seat - but without inflicting permanent physiological damage, therefore by eminence of the spiritual weight of the liver and by ignorance of its physiological functions? The liver is of course of paramount importance as we envision it today in its capacity to regenerate and maintain body physiology. We also know, above all, that the liver (i) has an important storage capacity, (ii) plays a primordial role as a metabolic powerhouse, (iii) has a secretory ability and endocrine gland attributes, (iv) is a predominant organ involved in the biodistribution of EVs [73, 74] and maybe vesicle-free miRNAs, and that its secretome is important [49, 75]. The liver can probably serve as a critical platform for the treatment of extracellular ncRNAs. We are tempted by the idea that the liver is capable of producing, absorbing, exchanging, storing and redistributing ncRNAs in transmission channels appropriate to the final targets, at least in part through HDL/LDL biogenesis.

Of course, the mechanisms mentioned, as well as others not mentioned, may exist in parallel to serve various physiological functions. There is, however, a lack of empirical evidence to support some or all of the possibilities mentioned, including the existence of mechanisms for the reception of miRNA vehicles or free miRNAs by certain cells, the arrangements undertaken with respect to stimulation signals, and the sorting and secretion in likely candidate tissues.

# Limitations and perspectives

Methodological difficulties remain the main obstacle when studying non-coding RNAs. Tracking a miRNA for example from its genesis or introduction to the fulfillment of its role in the destination is a very complicated and resource-intensive task. Still, we can be pleased with the recent large volume of critical experiments and empirical evidence which, by overcoming methodological difficulties, have shed light on some key logistical possibilities of RNAs circulating in the body. Nevertheless, efforts still need to be made to fully understand the extent of the involvement of these molecules, notably through the standardization of methodological processes. The technologies enabling the collection, extraction, detection, profiling and quantification of RNAs have experienced different accelerations in development, but are, for the most part, considered mature enough to allow a fairly accurate reporting of circulating RNAs. However, a methodological consensus would be of great assistance to this cause, which still suffers from the disparity of analytical results. In the present case, exploration of appropriate cell types, an accurate and consensual distribution picture, more rigorous carrier discrimination and in-vivo validations could eventually help unveil potentially revealing mechanisms. Also, experimentation on candidate tissues, using different delivery modes and upstream and downstream analysis, or blocking of transporters, could constitute initial experimental strategies. Finally, we are confident that the rapid trend in the development of information technologies and massive data learning and pattern interpretation will allow the elucidation of a number of scientific questions and will contribute to the confirmation or refutation of the NAcrin hypothesis.

The current perception of RNAs is causing a conflicting mental posture. Between consent on their crucial importance inside the cell and a perplexed looking towards those that are outside. As we cross the network of circulating RNAs, aren't we stumbling around the structures that produce and manage them? Far from making the NAcrins the scapegoat for all morbidities, we believe that the idea of their existence explains the phenomenon of extracellular RNAs, answers a number of questions in health and disease that have remained unresolved, and sheds light on the path towards unexplored mechanisms that govern physiology.